\documentclass[aps,prl,twocolumn,superscriptaddress,amsmath]{revtex4-1}

\usepackage{graphicx}

\begin{document}

\title{Observation of thermoelectric currents in high-field superconductor-ferromagnet tunnel junctions}

\author{S. Kolenda}
\affiliation{Institute of Nanotechnology, Karlsruhe Institute of Technology, Karlsruhe, Germany}
\author{M. J. Wolf}
\altaffiliation{Present addresss: Institute of Technical Physics, Karlsruhe Institute of Technology, Karlsruhe, Germany}
\affiliation{Institute of Nanotechnology, Karlsruhe Institute of Technology, Karlsruhe, Germany}
\author{D. Beckmann}
\email[e-mail address: ]{detlef.beckmann@kit.edu}
\affiliation{Institute of Nanotechnology, Karlsruhe Institute of Technology, Karlsruhe, Germany}

\date{\today}

\begin{abstract}
We report on the experimental observation of thermoelectric currents in superconductor-ferromagnet tunnel junctions in high magnetic fields. The thermoelectric signals are due to a spin-dependent lifting of particle-hole symmetry, and are found to be in excellent agreement with recent theoretical predictions. The maximum Seebeck coefficient inferred from the data is about $-100~\mathrm{\mu V/K}$, much larger than commonly found in metallic structures. Our results directly prove the coupling of spin and heat transport in high-field superconductors.
\end{abstract}



\maketitle

The coupling of electronic and thermal properties in nanoscale superconductor hybrid structures is an active field of current research due to applications as thermometers, microrefrigerators and particle detectors \cite{giazotto2006}. Recently, large thermoelectric effects were predicted to occur in superconductor-ferromagnet hybrid structures \cite{machon2013,*machon2014,ozaeta2014,kalenkov2014,*kalenkov2015}. Thermoelectric effects in metals are driven by broken symmetry between electron and hole carriers and are usually quite small, of the order of a few $\mathrm{\mu V/K}$ at room temperature and vanishing at low temperature.  While superconductors obey overall electron-hole symmetry, spin-splitting of the quasiparticle density of states breaks electron-hole symmetry for each spin band, and in conjunction with the spin-dependent conductance of a superconductor-ferromagnet tunnel junction leads to thermoelectric effects \cite{ozaeta2014}. The characteristic energy scale, the energy gap of the superconductor, is very small compared to the Fermi energy, and consequently, the thermoelectric effects are predicted to be large. Apart from being large, which makes them promising for applications, these thermoelectric effects are also interesting from a fundamental point of view. They lead to a coupling of spin and heat currents, which forms the basis of spin caloritronics \cite{bauer2012}. Thermally driven spin currents have recently been proposed \cite{silaev2015b} as an explanation for the experimentally observed long-range quasiparticle spin transport in superconductors \cite{huebler2012b,*wolf2013,*wolf2014c,quay2013}. Here, we report on the experimental observation of thermoelectric currents in high-field superconductor-ferromagnet tunnel junctions.

The current through a tunnel junction in the presence of a voltage $V$ and a temperature difference $\delta T$ across the junction can be conveniently described in the linear regime by
\begin{equation}
 I = gV+\eta \frac{\delta T}{\overline{T}}
   \label{eqn_Igeneral}
\end{equation}
where $g$ is the conductance, $\overline{T}$ is the average temperature, and $\eta$ describes the thermoelectric current. It is related to the Seebeck coefficient $S=-V/\delta T$ measured in an open circuit by $\eta=Sg\overline{T}$. Measuring the thermoelectric coefficient $\eta$ is the main purpose of this work.

\begin{figure}[bt]
\includegraphics[width=\columnwidth]{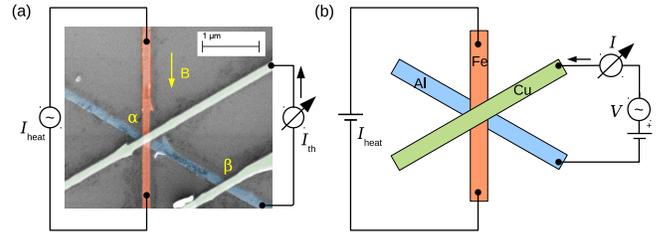}
\caption{\label{fig_sem}
(color online)
(a) False-color scanning electron microscopy image of a region of sample FIS1 with measurement configuration for the thermoelectric measurements.
(b) Sketch of the measurement configuration for conductance and heater calibration measurements.}
\end{figure}

Our samples were prepared by e-beam lithography and shadow evaporation technique. First a thin superconducting (S) aluminum strip of thickness $t_\mathrm{Al} \approx 20~\mathrm{nm}$ was evaporated, which then was oxidized {\em in situ} to form a thin insulating (I) tunnel barrier. After that ferromagnetic (F) iron ($t_\mathrm{Fe} \approx 15 -20~\mathrm{nm}$) and normal-metal (N) copper wires ($t_\mathrm{Cu} \approx 50~\mathrm{nm}$) were overlaid from different angles.  
In this manuscript we present and compare results from three samples, two samples with ferromagnet-superconductor junctions (FIS1 and FIS2), and a nonmagnetic reference sample (NIS). Figure~\ref{fig_sem}(a) shows a false-color scanning electron microscopy image of a region of sample FIS1. The central part of the structure is a 6-probe junction ($\alpha$) consisting of a superconductor-ferromagnet tunnel contact overlaid with an additional copper wire. This design allows us to measure the current through the tunnel junction while simultaneously passing a heater current along the iron wire to create a temperature difference across the junction. In addition, there is a 4-probe normal-metal junction ($\beta$), located at a distance $d \approx 1.5~\mathrm{\mu m}$ from $\alpha$. This junction is used for control experiments. Sample FIS2 has a slightly different layout (not shown) which avoids passing the heater current across the junction. As it turns out this had no effect on the results. The third sample (NIS) has the same layout as FIS1, with the iron wire replaced by a copper wire of a similar thickness. The measurement schemes are described below together with the results. Measurements were performed in a dilution refrigerator down to a base temperature of $T_0=50~\mathrm{mK}$, with an in-plane magnetic field $B$ applied parallel to the iron wire as indicated in Fig.~\ref{fig_sem}(a). We denote the temperature of the refrigerator by $T_0$ to distinguish it from the electronic temperature of the sample throughout this manuscript.

\begin{figure}[bt]
\includegraphics[width=\columnwidth]{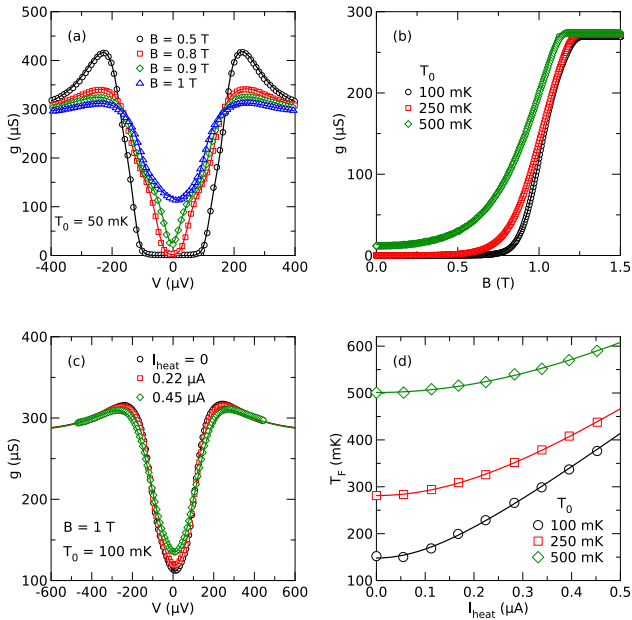}
\caption{\label{fig_contactA}
(color online)
(a) Differential conductance $g$ of junction $\alpha$ of sample FIS1 as a function of bias $V$ for different magnetic fields $B$.
(b) Zero-bias conductance $g$ as function of magnetic field $B$ for different base temperatures $T_0$.
(c) Differential conductance $g$ at fixed base temperature and magnetic field for different heater currents $I_\mathrm{heat}$.
(d) Temperature $T_\mathrm{F}$ of the ferromagnet as a function of $I_\mathrm{heat}$ for different $T_0$.}
\end{figure}

Before we describe the thermoelectric measurements, we first characterize the sample and calibrate the heater current.
In Figs.~\ref{fig_contactA}(a) and (b) we show the differential conductance $g$ of junction $\alpha$ of sample FIS1, measured by standard low-frequency ac technique in the configuration shown in Fig.~\ref{fig_sem}(b), without heater current. Figure \ref{fig_contactA}(a) shows $g$ as a function of bias voltage $V$ for different magnetic fields $B$. In high fields the Zeeman-splitting of the density of states is visible together with a broadening of the curves due to orbital pair breaking. To fit our data, we model the current using eqn. (2a) of Ref.~\onlinecite{ozaeta2014}. We assume that the superconductor is at temperature $T_\mathrm{S}=T$, and the ferromagnet is at temperature $T_\mathrm{F}=T+\delta T$, with voltage $V$ applied to the ferromagnet. This yields
\begin{multline}
 I(T,\delta T,V) = \frac{G_\mathrm{T}}{e}\int \left[N_0(E)+\frac{PN_z(E)}{2}\right]\\ 
\times\left[f_0(E-eV,T+\delta T)-f_0(E,T)\right]dE,
  \label{eqn_Ithermoelectric}
\end{multline}
where $G_\mathrm{T}$ is the normal-state junction conductance, $P$ is the spin polarization of the junction conductance, $e=-|e|$ is the charge of the electron, and $f_0$ is the Fermi function. Equation (\ref{eqn_Ithermoelectric}) includes conductance ($V\neq 0$, $\delta T=0$) and thermoelectric currents ($V=0$, $\delta T\neq0$) on an equal footing.
The density of states factors are $N_0=(N_{+}+N_{-})/2$ and $N_z=N_{+}-N_{-}$, where $N_\pm$ are the densities of states for the two spin projections in the superconductor. $N_+$ and $N_-$ were obtained from the standard model of high-field tunneling \cite{maki1964b,meservey1975}, including the effect of Zeeman splitting, orbital pair breaking and spin-orbit scattering. The factor $PN_z$ is odd in energy, and gives rise to the observed thermoelectric effect. It is non-zero only in the presence of a Zeeman splitting of the density of states in combination with a spin-polarized junction conductance.
Fits of our data based on eqn. (\ref{eqn_Ithermoelectric}) are shown as lines in Fig.~\ref{fig_contactA}(a). From these fits we extract the spin polarization $P\approx 0.08$. Due to the finite spin polarization $P$, there is a small but visible asymmetry of the conductance at high fields \cite{tedrow1971,*meservey1994}, with higher conductance for electrons (negative bias) than holes (positive bias).

In Fig.~\ref{fig_contactA}(b) we show $g$ at zero bias as a function of the magnetic field $B$ for three base temperatures $T_0$, together with fits to our model. For these fits, the self-consistent pair potential $\Delta$, orbital pair-breaking strength $\alpha_\mathrm{orb}$ and effective Zeeman splitting were calculated as a function of the magnetic field using Ref.~\onlinecite{alexander1985}, including the effect of Fermi liquid renormalization on the effective Zeeman splitting. The latter, with the Fermi liquid parameter $G^0=0.3$ taken from literature \cite{tedrow1984}, was found to improve the fits close to the critical field. The fits of the conductance are used to establish the necessary sample parameters for modeling the thermoelectric results shown below.

We now turn to the heater calibration. In Fig.~\ref{fig_contactA}(c) we show the differential conductance $g$ as a function of bias $V$ at fixed temperature $T_0 = 100~\mathrm{mK}$ and magnetic field $B= 1~\mathrm{T}$ for different dc heater currents $I_\mathrm{heat}$ (see Fig.~\ref{fig_sem}(b) for the schematics). Upon increasing $I_\mathrm{heat}$, the conductance visibly broadens due to increasing temperature $T_\mathrm{F}$ of the ferromagnet (the thermal broadening depends on $T_\mathrm{F}$ only and is independent of $T_\mathrm{S}$). From fits of the data we extract $T_\mathrm{F}$ as a function of $I_\mathrm{heat}$, shown in Fig.~\ref{fig_contactA}(d) for three different base temperatures $T_0$. 

To describe the heating of the junction as a function of current, we assume quasiequilibrium with negligible electron-phonon scattering, as appropriate for mesoscopic metal wires of about $10~\mathrm{\mu m}$ length at sub-Kelvin temperatures \cite{giazotto2006}. In this case, the temperature at the junction (in the middle of the heater wire) is given by
\begin{equation}
 T_\mathrm{F}=\sqrt{T^2+\frac{I_\mathrm{heat}^2R_\mathrm{heat}^2}{4L_0}},
   \label{eqn_heatmodel}
\end{equation}
where $R_\mathrm{heat}$ is the resistance of the heater wire, $L_0=\pi^2k_\mathrm{B}^2/3e^2$ is the Lorenz number, and $T$ is the electronic base temperature in the absence of heating. In Fig.~\ref{fig_contactA}(d) we show fits of our data using eqn.~(\ref{eqn_heatmodel}). As can be seen, the fits show a good agreement with the data, and we use the fits for temperature calibration for the thermoelectric measurements described below. The electronic base temperature $T$ is slightly increased over the cryostat base temperature $T_0$ at low temperatures, probably due to incomplete filtering of thermal noise from higher temperature stages of the cryostat. The fits yield $R_\mathrm{heat} \approx 230~\mathrm{\Omega}$, smaller than the 2-probe resistance of the iron wire, $R_\mathrm{Fe} = 1350~\mathrm{\Omega}$. We attribute this to the fact that the thick copper wire acts as a cooling fin.

\begin{figure}[bt]
\includegraphics[width=\columnwidth]{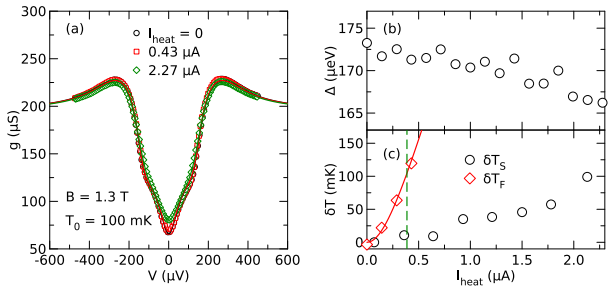}
\caption{\label{fig_contactB}
(color online)
(a) Differential conductance $g$ of contact $\beta$ of sample FIS2 at fixed magnetic field $B$ and temperature $T_0$ for different heater currents $I_\mathrm{heat}$.
(b) pair potential $\Delta$ as a function of $I_\mathrm{heat}$ obtained by fitting the differential conductance shown in (a).
(c) increase $\delta T_\mathrm{S}$ of the temperature of the superconductor as a function of $I_\mathrm{heat}$ extracted by inverting $\Delta(T_\mathrm{S})$, together with $\delta T_\mathrm{F}$.}
\end{figure}

Superconductors are poor heat conductors at low temperature, and can therefore be easily heated. To check the impact of the heater current on the temperature $T_\mathrm{S}$ of the superconductor we performed additional control experiments on junction $\beta$.  The quasiparticle energy relaxation length is typically a few $10~\mathrm{\mu m}$ in aluminum at low temperatures \cite{arutyunov2011}, and we assume $T_\mathrm{S}$ to be nearly the same at contacts $\alpha$ and $\beta$.  An increase of $T_\mathrm{S}$ affects the differential conductance $g$ only indirectly by a small reduction of the pair potential $\Delta$. In Fig.~\ref{fig_contactB} we show data of sample FIS2, for which we made the most detailed measurements of $T_\mathrm{S}$. Figure~\ref{fig_contactB}(a) shows the differential conductance $g$ for different heater currents $I_\mathrm{heat}$ at fixed magnetic field $B = 1.3~\mathrm{T}$ and temperature $T_0 = 100~\mathrm{mK}$. As can be seen, there is almost no change at small currents, and at larger currents, the gap is slightly reduced. The pair potential $\Delta$ obtained by fitting the data is plotted in Fig.~\ref{fig_contactB}(b) as a function of the heater current. The change of $\Delta$ is of the order of a few $\mathrm{\mu eV}$ for the largest applied current, with considerable scatter on this small scale. To reduce scatter, we averaged data for two adjacent points and then inverted the self-consistent relation $\Delta(T+\delta T_\mathrm{S},B)$ to obtain $\delta T_\mathrm{S}$ (assuming $\delta T_\mathrm{S}=0$ for the first point). In Fig.~\ref{fig_contactB}(c) the resulting $\delta T_\mathrm{S}$ is shown as a function of $I_\mathrm{heat}$, together with $\delta T_\mathrm{F}$. For thermal bias $\delta T_\mathrm{F} < 100~\mathrm{mK}$ used in the thermoelectric experiments (indicated by the dashed line), we find $\delta T_\mathrm{S} \lesssim 20~\mathrm{mK}$, much smaller than $\delta T_\mathrm{F}$. We conclude that most of the thermal bias actually drops across the junction.

\begin{figure}[bt]
\includegraphics[width=\columnwidth]{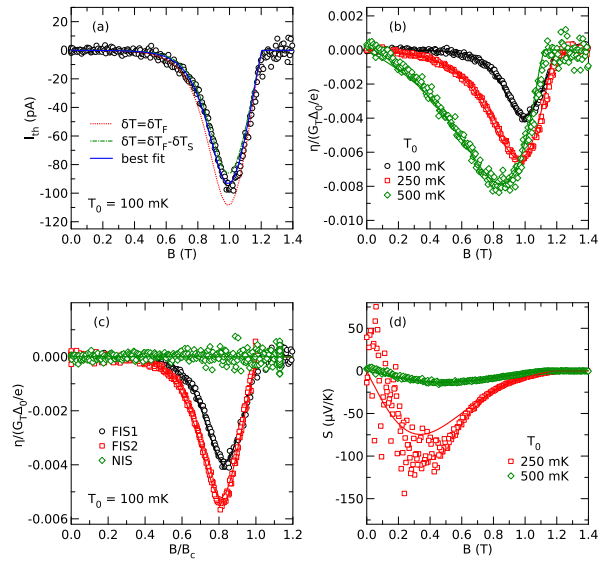}
\caption{\label{fig_thermoelectric}
(color online) 
(a) Thermoelectric current $I_\mathrm{th}$ as a function of magnetic field $B$ at base temperature $T_0 = 100~\mathrm{mK}$, together with theoretical calculations (see text).
(b) thermoelectric coefficient $\eta$ of sample FIS1 for different temperatures $T_0$. 
(c) thermoelectric coefficient $\eta$ at $T_0=100~\mathrm{mK}$ for all three samples. 
(d) Seebeck coefficient $S$ as a function of $B$ calculated as described in the text.}
\end{figure}

Now we will focus on the thermoelectric measurements. For these measurements, we applied an ac heater current, as shown schematically  in Fig.~\ref{fig_sem}(a). Since the heating power is proportional to $I^2_\mathrm{heat}$, the thermal excitation appears on the second harmonic of the signal, and we detected the second harmonic of the resulting current $I_\mathrm{th}$ flowing into the superconductor. The heater current amplitude was typically chosen to yield a peak-to-peak thermal amplitude $\delta T_\mathrm{F} =100~\mathrm{mK}$. In Fig.~\ref{fig_thermoelectric}(a) we show the peak-to-peak amplitude of the thermoelectric current $I_\mathrm{th}$ as function of the magnetic field $B$ at base temperature $T_0 = 100~\mathrm{mK}$ for sample FIS1. As expected, there is no signal at zero field, {\em i.e.}, in the absence of Zeeman splitting. With increasing field a negative signal develops above $B = 0.5~\mathrm{T}$ and reaches a broad maximum around $B = 1~\mathrm{T}$, then going back to zero as $B$ approaches the critical field. The thermoelectric current is negative, as expected for our junction with higher conductance for electrons than holes, as seen in Fig.~\ref{fig_contactA}(a).
We also show theoretical calculations using eqn.~(\ref{eqn_Ithermoelectric}) with the same parameters as for the fits of the conductance shown in Fig.~\ref{fig_contactA}(b). For the amplitude $\delta T$ we chose two assumptions: The dashed line is calculated using $\delta T=\delta T_\mathrm{F}=100~\mathrm{mK}$ obtained from the calibration fit to eqn.~(\ref{eqn_heatmodel}), whereas the dash-dotted line is calculated using $\delta T=\delta T_\mathrm{F}-\delta T_\mathrm{S}\approx 80~\mathrm{mK}$. The data lie in between both curves, and the best fit (solid line) is obtained for $\delta T = 86~\mathrm{mK}$. 
In Fig.~\ref{fig_thermoelectric}(b) we show the thermoelectric coefficient $\eta = I_\mathrm{th}\overline{T}/\delta T$ for sample FIS1 as a function of the magnetic field $B$ for different base temperatures $T_0$. Here $\overline{T}=T+\delta T/2$, and we used $\delta T$ from the best fits. For increasing temperatures we see that both the magnitude and broadening of $\eta$ increase, as expected from theory \cite{ozaeta2014}.
Figure~\ref{fig_thermoelectric}(c) shows $\eta$ as a function of the normalized field $B/B_\mathrm{c}$ for all samples at $T_0 = 100~\mathrm{mK}$. The data are similar for both ferromagnetic samples FIS1 and FIS2. There is no signal for the non-magnetic reference sample NIS, directly proving the spin-dependent origin of the effect. From the data and fits in Figs.~\ref{fig_thermoelectric}(a)-(c), we conclude that the data are in excellent qualitative and quantitative agreement with the theoretical predictions \cite{ozaeta2014}. 
 
In Fig.~\ref{fig_thermoelectric}(d) we finally show the Seebeck coefficient $S = \eta/(g\overline{T})$ inferred from our measurements as a function of the applied magnetic field $B$ for sample FIS1. The data and fits of $S$ were both calculated from the data and fits of $\eta$ and $g$ shown in Figs.~\ref{fig_thermoelectric}(b) and \ref{fig_contactA}(b). As can be seen, $S$ shows a similar field dependence as $\eta$ itself, however with a faster increase and maximum at smaller fields. The Seebeck coefficient is larger at lower temperature due to the freeze-out of quasiparticle conductance. As a result, the scatter is also amplified, and the data for $T_0=100~\mathrm{mK}$ are unreliable and omitted from the plot. For $T_0 = 250~\mathrm{mK}$, the maximum value is $S\approx -100~\mathrm{\mu V/K}$.

In conclusion, we have measured thermoelectric currents in superconductor-ferromagnet tunnel junctions at high magnetic fields. The results are in excellent agreement with recent theoretical predictions. The Seebeck coefficients inferred from the data can be as large as $100~\mathrm{\mu V/K}$, much larger than usually found in metal structures, despite the fact that the spin polarization of our tunnel junctions is relatively small. With larger spin polarization, e.g., using spin-filter junctions \cite{hao1990}, Seebeck coefficients exceeding $1~\mathrm{mV/K}$ seem feasible. Our results directly prove the coupling of spin and heat transport in high-field superconductors.


\bibliography{thermoelectric.bib}

\end{document}